\documentclass[twocolumn,3p,article]{elsarticle}
\usepackage{lineno} 

\usepackage{amsmath}
\usepackage{breqn}
\usepackage{graphicx}
\biboptions{super,square,sort&compress,comma,numbers}
\journal{J.~Comp.~Chem.}
\begin{document}
\begin{frontmatter}
%
\title{CPMD/GULP QM/MM Interface for Modeling Periodic Solids: Implementation and its Application in the Study of Y$-$Zeolite Supported Rh$_n$ Clusters}
%
\author[mymainaddress]{Sudhir K. Sahoo}
\author[mymainaddress]{Nisanth N. Nair\corref{mycorrespondingauthor}}
\address[mymainaddress]{Department of Chemistry, Indian Institute of Technology, Kanpur, 208016, India}
\cortext[mycorrespondingauthor]{Corresponding author}
\ead{nnair@iitk.ac.in}
\begin{abstract}
We report here the development of 
hybrid quantum mechanics/molecular mechanics (QM/MM) interface between the 
 plane--wave density functional theory based CPMD code and the empirical force--field based GULP code for
 modeling periodic solids and surfaces.
The hybrid QM/MM interface is based on the electrostatic coupling between QM and MM regions.
The interface is designed for carrying out full relaxation of all the QM and MM atoms during geometry optimizations and molecular dynamics simulations, including the boundary atoms.
Both Born--Oppenheimer and Car--Parrinello molecular dynamics schemes are enabled for the QM part during the QM/MM calculations.
This interface has the advantage of parallelization of both the programs such that the QM and MM force evaluations can be carried out
in parallel in order to model large systems.
The interface program is first validated for total energy conservation and parallel scaling performance is benchmarked. 
Oxygen vacancy in $\alpha$--cristobalite is then studied in detail and the results are compared with a fully QM calculation and experimental data.
%
%
%
Subsequently, we use our implementation to investigate the structure of rhodium cluster (Rh$_n$; $n$=2 to 6) formed from Rh(C$_2$H$_4$)$_2$ complex 
adsorbed within a cavity of Y--zeolite in a reducible atmosphere of H$_2$ gas.
%
%
\end{abstract}
\end{frontmatter}
%
%
\section{Introduction}
%
Quantum mechanics (QM) potential based molecular dynamics (MD) simulations 
(known as {\em ab initio} MD) of industrially important catalytic reactions 
is challenging due to the length--scale of the system and the time--scale 
of the processes that need to be addressed. 
A pragmatic way to bridge the length--scales (for such problems) is by using  
hybrid quantum mechanics/molecular mechanics (QM/MM)~\cite{Warshel_Levitt_1976} 
methods.~\cite{Monard_acr_1999,Bernstein_Rep.Prog.Phys_2007,Maseras_Dalton_Trans_2008,Thiel_angewie_2009,Victor_WIRES_2011,Lin_ms_2015,Brunk_cr_2015} 
In a QM/MM calculation, a small portion of the catalytic system (i.e. the active site) is treated by a QM potential, 
while the rest of the large portion of the system is treated by a computationally cheap empirical 
molecular mechanics (MM) potential.
%
A few hundreds of pico--second long MD simulation for $\sim$10$^5$ MM atoms and $\sim10^2$ 
QM atoms in a QM/MM simulation is feasible today. 
In order to bridge the differences in time scales of such simulations 
($\sim10^2$~ps) and that of a catalytic processes (seconds to hours), 
enhanced sampling MD techniques, such as metadynamics~\cite{Laio_pnas_2002} 
have to be used.
Thus to address complex catalytic processes computationally, 
a QM/MM interface program which can carry out enhanced sampling 
MD simulations of chemical reactions would be ideal.

A particularly interesting catalytic system is the supported catalysts, where 
metallic clusters adsorbed on a solid support is the active site.~\cite{Anderson_sup_met_catl} 
%
Zeolite--supported metal clusters are potential catalysts for various chemical reactions.~\cite{Portugal_aca_g_2000,Guzman_Dalton_Trans_2003,Althoff_catal_rev_2008,Roth_catt_2012} 
Computational modeling of zeolites using a (fully) QM potential is extremely computationally demanding 
due to large number of atoms in its unit cell and its large volume.
On the other hand, a QM/MM method is optimal to model such systems. 
QM/MM approaches have been developed to study zeolite and similar silica 
systems.~\cite{Clark_jmm_1999,Kramer_JPCB_1999,Sauer_JCC_JCC5,Sherwood_Theochem_2003,Sulimov_prb_2002,Vladimir_JPCB_2003,Cheng_QC_2003,Zipoli_jcp_2006,Zimmerman_JCTC_1011}
%

%
A key component of the QM/MM method is the QM--MM electrostatic interaction. 
In a plane--wave density functional theory (DFT) based approach employing an electronic embedding scheme, interaction between
electron density and MM point charges has to be accounted in the Hamiltonian.
Here QM density is solved in the presence of the external field of MM charges and 
a total energy conserving QM/MM MD Hamiltonian can be formulated.~\cite{Laio_jcp_2002} 
A computationally efficient dual space--grid based QM/MM implementation 
for plane--wave basis was suggested by Yarne et~al.~\cite{Yarne_jcp_2001} 
and similar approach has been also implemented for a mixed Gaussian/plane--wave basis set.~\cite{Laino_jctc_2005,Laino_jctc_2006} 
In these implementations, a modified Coulomb kernel function is used to avoid over-polarization of electronic density
(or electron spill out) due to the MM charges, which is severe in the case of plane--wave basis.
A low MM point charge based MZHB force--field has been proposed to avert electron spill out for 
QM/MM applications in silica.~\cite{Sudhir_JCC_2015} 
Dangling bonds at the QM/MM boundary can be either treated by pseudopotentials~\cite{Zhang_jcp_1999} 
or by introducing capping hydrogen atoms.~\cite{Singh_jcc_1986} 
%

%
Here we report the development of a QM/MM interface code for treating periodic solids 
in order to study complex catalytic processes using MD techniques. 
We choose CPMD~\cite{cpmd1} as the QM code mainly due to its efficient implementation of 
Car--Parrinello molecular dynamics (MD) and 
methods such as metadynamics~\cite{Laio_pnas_2002} 
CPMD can carry out pseudopotential based plane--wave DFT computations.
As the MM code, we choose the GULP~\cite{Gale_gulp-ref_ms} program, mainly because of its simplicity 
in usage and it has implementations of variety of popular empirical force--fields for treating periodic solids.
Moreover, both CPMD and GULP have parallel implementations using Message Passing Interface (MPI), which is advantageous for treating
large systems.
A number of QM/MM interfaces for treating bimolecular systems are available for CPMD, such as the
CPMD/EGO by Eichinger et~al.~\cite{Eichinger_jcp_1999}, CPMD/GROMOS by Laio et~al.,~\cite{Laio_jcp_2002} 
CPMD/Gromacs interface by Biswas and Gogonea~\cite{Biswas_jcp_2005} and 
CPMD/Iphigenie by Schw{\"o}rer et~al.~\cite{Schworer_jcp_2013}.
However, the available QM/MM interfaces do not allow one to use variety of empirical potentials 
for treating periodic solids and has motivated us to develop the CPMD/GULP QM/MM interface.

In this work, we apply the developed CPMD/GULP QM/MM interface for studying zeolite supported Rh$_n$ clusters.
Recently B.~C.~Gates and coworkers~\cite{Liang_jpcc_2008,Serna_PCCP_2014}
have reported the synthesis and characterization of small Rh clusters
from mononuclear Rh complex supported in Y--zeolite.
The formation of Rh cluster depends on the temperature and on the nature of the 
ligands and the support.~\cite{Serna_PCCP_2014}
Rh--Rh distance measured using the extended X-ray absorption fine structure (EXAFS) technique
and infrared (IR) spectrum~\cite{Serna_JCatal_2013} reveals the presence
of H atom impurity in the Rh$_n$ cluster.
These Rh complexes and clusters are active catalyst for dimerization and hydrogenation reactions of ethene
and their catalytic activity depends on the nuclearity of the
catalyst.~\cite{Serna_jacs_2011}
We have investigated the thermodynamically most stable protonation states of Rh$_n$ clusters, $n=2-6$, on Y--zeolite
as a function of temperature and pressure.
To compute the thermodynamic stability of a protonated form of a cluster, we considered its formation
from Rh(C$_2$H$_4$)$_2$ complex in the presence of H$_2$ gas. 
R{\"o}sch and co--workers have addressed supported Rh clusters using QM/MM and DFT calculations.
The reverse hydrogen spillover reactions on Rh$_6$ and other late--transition metal clusters
supported by zeolites have been studied in detail.~\cite{Vayssilov_angew_2003,Vayssilov_pccp_2012} 
%
Benchmark studies of QM/MM calculations have been reported for 
C--C coupling and hydrogenation reactions involving zeolite--supported 
[Rh(C$_2$H$_4$)$_2$(H$_2$)]$^+$.~\cite{Dinda_jpcc_2014}
Structure and properties of protonated Rh$_n$ clusters in Y--zeolite
were also investigated using periodic DFT.~\cite{Markova_jpcc_2015}

In this paper, we first present the technical and computational details of the CPMD/GULP QM/MM 
implementation, followed by results. 
Test on the accuracy of the implementation is carried out by
verifying the total energy conservation in a microcanonical ensemble MD simulation of Y--zeolite.
Parallel efficiency of both QM and MM parts in our QM/MM implementation is then demonstrated. 
We study the neutral oxygen vacancy in $\alpha$--cristobalite and compare
the defect formation energy, and electronic, structural and dynamic properties from 
a canonical ensemble QM/MM MD simulation with a fully QM simulation and with available experimental data.
Finally, we investigate the free energy of formation of Rh$_n$H$_m$ ($n$=2 to 6) clusters supported on Y$-$zeolite,
at various temperatures.
 Properties of stable Rh$_n$H$_m$/Y clusters at 300~K is then obtained by QM/MM MD simulations.

\section{Methods}
\subsection{QM/MM Coupling}
%
QM/MM potential energy of a system is given by,
\begin{equation} \label{ETOT}
E_{\rm PE}=E_{\rm QM}+E_{\rm MM}+E_{\rm QM/MM} \enspace ,
\end{equation}
where $E_{\rm QM}$, $E_{\rm MM}$ and $E_{\rm QM/MM}$ are the energies of QM subsystem, MM subsystem and  interaction energy between QM and MM subsystems, respectively.
%
%
 $E_{\rm QM/MM}$ is  composed of electrostatic and van der Waals interactions, where the electrostatic interaction is computed between an MM point charge and QM electron density following Ref.~\cite{Laio_jcp_2002} as
\begin{equation} \label{E_EL}
E_{\rm EL}=\sum_{I\varepsilon \rm MM } q_{I}\int d{\rm \bf r} \rho({\rm {\bf r}}) \frac{r_{\mathrm cI}^{4}-r^{4}}{r_{\mathrm cI}^5-r^{5}}
\end{equation}
where a modified Coulomb kernel is used to avoid electron spill out problems.
%
%
Here,  $q_I$ is the point charge of an MM atom $I$, $\rho({\rm {\bf r}})$ is the electron density at a real space grid ${\bf r}$, $r_{\mathrm cI}$ is the covalent
 radius parameter for an atom $I$, and $r$ is the distance between atom $I$ and the grid ${\bf r}$ representing $\rho({\rm {\bf r}})$.
%
A more efficient approach to carry out electrostatic coupling will be to employ the dual space grid based coupling proposed in Ref.~\cite{Yarne_jcp_2001}, 
and later by Laino~et al.~\cite{Laino_jctc_2005,Laino_jctc_2006}, 
and will be implemented and tested in the future.

When the QM/MM boundary passes through a covalent bond special care has been taken.
In the presented work, we follow a link atom scheme where H atoms are used to saturate the dangling bonds of a QM atom at the QM/MM boundary. 
The electrostatic interactions computed using~\eref{E_EL} are excluded between bonded pair of atoms of QM and MM regions by giving an exclusion list.
The electrostatic interaction between bonded pair of atoms spanning across the QM and MM regions are computed as per the MM potential.
The MM atoms in this exclusion list is interacting with the rest of the QM atoms through
dynamically generated electrostatic potential (D--RESP) derived charges.~\cite{Laio_jpcb_2002}

The QM/MM Lagrangian in the framework of Car--Parrinello scheme is given by, 
\begin{dmath} 
\label{CP_Lg}
\mathcal L_{\rm CP/QMMM}=\sum_{I}\frac{1}{2}M_{I}\dot{\bf{R}}_{I}^2+\sum_{i} \frac{1}{2}\mu \left \langle{\dot{\phi_i}}|{\dot{\phi_i}}\right\rangle-E_{\rm {KS}}-E_{\rm {MM}}-E_{\rm {QM/MM}}
+\sum_{i,j}\Lambda _{ij}\left(\left\langle \phi_i\mid \phi_j\right\rangle-\delta_{i,j}\right)
\end{dmath}
Here the first and the second terms are  the kinetic energy of nucleus and fictitious kinetic energy of orbitals.
$E_{\rm {KS}}$ is the Kohn--Sham energy which is identical to $E_{\rm QM}$ in \eref{ETOT}.
The last term ensures that the orthonormality constraints are fulfilled during the time evolution of wavefunctions.
\subsection{Implementation Details}
%
%
%
Our QM/MM implementation that combines both CPMD and the GULP codes, makes use of the efficient parallel implementation of both
these programs using MPI.
For this purpose, two MPI communication groups are formed within the CPMD program to handle communications within the CPMD and the GULP programs independently.
The communication between these two MPI groups are only to exchange the coordinates and the forces at every MD step, and thus the
communication overhead is usually insignificant.
Both the programs run in parallel during the force--evaluation, which is crucial to address large systems of interest.
The CPMD code is the main MD driver which allows the user to take advantages of different existing MD features of the CPMD package, in particular,
Car--Parrinello MD.
%
%
%
%
\subsection{Benchmark calculations using FAU zeolite} 
\label{m:bench:1}

To validate the code, a $1{\times}1{\times}1$ supercell of pure silica FAU zeolite (Si$_{192}$O$_{384}$) was taken.
We performed $NVE$ ensemble simulation using the QM/MM Hamiltonian, where a 
Si$_2$O$_7$ unit  
was treated within the QM region and the rest in the MM region.
The 6 oxygen atoms at the QM/MM boundary were capped by H atoms.
Free boundary conditions were used for the QM supercell (since the basis functions are plane waves) 
to obtain non--periodic electronic density.
QM subsystem was taken in a cubic supercell of side 13.22~{\AA}.
%
A cubic periodic supercell of size 24.50~{\AA} was used for the whole system,
which was initially optimized using the MM force-field.
For further testing the scalability of both CPMD and GULP, a $2{\times}2{\times}2$ supercell of FAU zeolite (Si$_{1536}$O$_{3072}$) was taken. 
The QM region contained a T6 site (Si$_6$O$_{18}$) and 12 capping H atoms for terminal O atoms.
The QM system was taken in a cubic supercell of edge length 17.46~{\AA},  
and the cubic periodic supercell with side 48.96~{\AA} was used for the whole system.

The MM part was treated using the low--point charge MZHB potential for SiO$_2$ from Ref.~\cite{Sudhir_JCC_2015}
This low--point charge force--field was necessary to avoid 
 over polarization of the electron density due to large MM point charges, especially
 because wavefunction is described using plane wave basis set.
A 30~Ry plane wave cutoff was used to expand wavefunction,  
and  ultrasoft pseudopotentials~\cite{Vanderbilt_prb_1990} were used to treat core electrons.
%
All the calculations were closed shell and were using the PBE~\cite{Perdew_prl_1996} exchange correlation functional.


%
In QM/MM $NVE$ simulations, all the QM and MM atoms were let free to move, but the capping hydrogens were
constrained to move in the direction of the Si--O bond along the QM/MM boundary.
MD simulations of the QM part were carried out using the Car--Parrinello scheme,~\cite{Parrinello_prl_1985} 
 while both the QM and the MM atoms were propagated
 by a timestep of 4~a.u.
The fictitious orbital mass in the Car--Parrinello dynamics was set to 600 a.u. and hydrogen atom mass (for the capping atoms) 
was replaced with deuterium.
For $NVE$ simulations, no thermostats were coupled to the orbital degrees of freedom.

\subsection{Neutral oxygen vacancy in $\alpha$--Cristobalite}
\label{m:bench:2}
The neutral oxygen vacancy in $\alpha$--Cristobalite has been studied to benchmark
the performance of the CPMD/GULP QM/MM code.
The defect formation energy in this system computed using periodic DFT (full QM) is compared with that computed
using the CPMD/GULP QM/MM code.
%
%
The vacancy formation energy, $\Delta E_{\rm f}$,  was computed as,
\begin{dmath}
\label{vcf}
\Delta E_{\rm f}=\frac{1}{2}\left [ E({\rm O_2})+E_{\rm diss}({\rm O_2}) \right ] +E({\rm SiO_{2-\it x}})-E({\rm SiO_2})
\end{dmath}
Here, $E({\rm O_2})$, $E_{\rm diss}({\rm O_2})$, $E({\rm SiO_{2-\it x}})$, and, $E({\rm SiO_2})$
are energy of O$_2$ molecule (in the triplet electronic ground state), dissociation energy of O$_2$ molecule, 
energy of bulk silica with oxygen vacancy, and energy of pure bulk silica, respectively.
 $E({\rm O_2})$ was computed from QM calculation whereas $E({\rm SiO_{2-\it x}})$ and $E({\rm SiO_2})$ were computed either by QM or QM/MM calculations.
 $E_{\rm diss}({\rm O_2})$ was taken from experimental data (5.16 eV).~\cite{CRC_Handbook} 

\begin{figure}
\begin{center}
\includegraphics[width=1.00\linewidth]{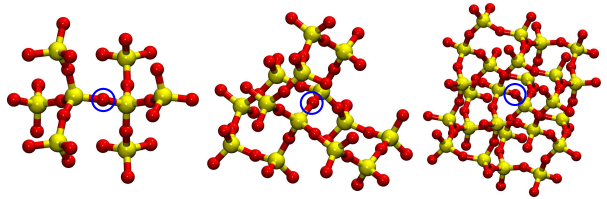}
\caption[]{
\label{QMMM_Sites}
Structure of different QM subsystems (from the left 8T, 14T, and, 26T) in the QM/MM calculations are shown.
Atom color code: Si--yellow and O--red. 
Oxygen vacancy was created by removing the highlighted oxygen atom (blue circle) from the lattice.}
\end{center}
\end{figure}

The initial lattice parameters of the system were taken from Ref.~\cite{Dollase_1965}, which were further optimized using the MM force--field.\cite{Sudhir_JCC_2015} 
For the fully QM periodic DFT calculation, three different supercells, 
2$\times$2$\times$2 (Si$_{32}$O$_{64}$) , 
3$\times$3$\times$2 (Si$_{72}$O$_{144}$) , 
and 3$\times$3$\times$3 (Si$_{108}$O$_{216}$) 
were taken, while in the QM/MM simulations a supercell of size 
{9$\times$9$\times$9} (Si$_{2916}$O$_{5832}$) 
was considered.
%
%
We performed three set of QM/MM calculations with different sizes of the QM--subsystems: 8T (Si$_8$O$_{25}$), 
14T (Si$_{14}$O$_{40}$), 
and 
26T (Si$_{26}$O$_{93}$), 
where T stands for a (shared) SiO$_4$ tetrahedral site; see~\fref{QMMM_Sites}.
Oxygen vacancy was created by removing one of the oxygen atoms (in the QM region) from its lattice position (as shown in Fig.~\ref{QMMM_Sites}).

We employed simulated annealing of both nuclei and orbitals within the Car--Parrinello scheme to optimize the structure, followed
by a quasi-Newton based optimization for nuclei and conjugate gradient minimization for orbitals, whenever necessary.
$NVT$ ensemble simulations were carried out at 300 K using  Nos{\'e}--Hoover Chain (NHC) thermostats.~\cite{Martyna_jcp_1992} 
%
The whole QM and the MM atoms were relaxed during all the MD simulations runs. 
All the other computational details are identical to that in Section~\ref{m:bench:1}. 

\subsection{Study of Zeolite Supported Rh$_n$ ($n=2-6$) Clusters}
\label{m:Rh:zeolite}
As an application of our implementation, we study the structure of protonated Rh$_n$, $n=2-6$ in HY--zeolites.
Cluster formation is investigated at a range of temperature and pressure by employing {\it ab--initio} 
thermodynamics techniques.~\cite{Reuter_prb_2001,Meyer_prb_2004} 
Based on the experiments~\cite{Liang_jpcc_2008} 
where cluster formation has been observed, 
we modelled the cluster formation reaction for $n>2$ as,
\begin{dmath}
{\rm Rh}_n{\rm H}_m/{\rm Y}+{\rm Rh}({\rm C}_2{\rm H}_4)_2/{\rm Y}+x{\rm H}_2 \rightarrow {\rm Rh}_{n+1}{\rm H}_{m'}/{\rm Y}+{\rm HY}+2{\rm C_2H_6}
\label{chem_rxn_1}
\end{dmath}
where $m^\prime=m+2x-5$, 
and odd number of $m$ and $m^\prime$ values are only considered in this study
assuming that H$_2$ adsorption in Rh$_n$H cluster (and not the bare Rh$_n$ cluster) due to
reverse spillover reaction. 
Protonation of the Y zeolite is incorporated by the formation of HY during this reaction. 
The maximum value of $m$ and $m^\prime$ (and thus $x$) are determined by the maximum number of hydrogen atoms that can 
be adsorbed on a given cluster (for $x>0$).

%
%
%
The maximum number of H atom in the Rh$_n$ cluster is $2n+1$ as the optimum H/Rh ratio is 2.~\cite{Petkov_jpcc_2010}
Thus we considered 19 different cluster formation chemical reactions and are listed in the 
Supporting Information.

Free energy of these cluster formation reactions were computed as 
\begin{multline} \label{G_cf}
\Delta G_{\rm cf}(T,p)=G({\rm Rh}_{n+1}{\rm H}_{m'}/\mathrm Y)+G({\rm HY})\\ 
+2\mu_{\rm C_2H_6}-G({\rm Rh}_n{\rm H}_m/{\rm Y}) \\
-G({\rm Rh}({\rm C}_2{\rm H}_4)_2/{\rm Y})-x\mu_{{\rm H}_2}
\end{multline}
The enthalpy of zeolite supported metal complex/clusters is approximated by 
the potential energy ($E$) of the system as the effect due to the change their volume 
is negligible compared to other terms in the equation for the range $p$ and $T$ considered here.
%
%
%
We rewrite \eref{G_cf} as,
\begin{multline}
\label{G_cf_2}
\Delta G_{\rm cf}(T,p)=\Delta E_{\rm cf}-T\Delta S_{\rm cf}(T) \\ 
      +2\Delta \mu_{{\rm C}_2{\rm H}_6}\left(T,p_{{\rm C}_2{\rm H}_6}\right) \\ 
     -x\Delta \mu_{{\rm H}_2}\left(T,p_{{\rm H}_2}\right)
\end{multline}
%
where
\begin{multline} 
\label{E_cf}
\Delta E_{\rm cf}=E({\rm Rh}_{n+1}{\rm H}_{m'}/Y)+E({\rm HY}) \\ 
      +2E_{\rm C_2H_6}-E({\rm Rh}_n{\rm H}_m/{\rm Y}) \\ 
      -E({\rm Rh}({\rm C}_2{\rm H}_4)_2/{\rm Y})-xE_{{\rm H}_2} \enspace , 
\end{multline} 
%
%
\begin{multline} \label{mu_cf}
\Delta \mu(T,p)=\Delta H(T,p^0)-T\Delta S(T,p^0) \\ 
              +k_{\rm B}T \ln\left (\frac{p}{p^0}\right) \enspace ,
\end{multline}
%
%
and 
\begin{multline} 
\label{dS_cf}
\Delta S_{\rm cf}(T)=S_{{\rm Rh}_{n+1}{\rm H}_{m'}/{\rm Y}}(T)+S_{\rm HY}(T) \\ 
      -S_{{\rm Rh}_n{\rm H}_m/{\rm Y}}(T)-S_{{\rm Rh}({\rm C}_2{\rm H}_4)_2/{\rm Y}}(T) \enspace .
\end{multline}
Here $k_{\rm B}$ is the Boltzmann constant, and $p^0$ is the standard pressure. Partial pressure of a gaseous component $i$ is indicated as $p_i$. 
$E_{\rm C_2H_6}$ and $E_{\rm H_2}$ in \eref{E_cf} were computed using QM calculations. The individual energy terms in \eref{E_cf} corresponds to the
most stable structure of the corresponding species among various configurations of the system explored in simulated annealing optimizations.
In \eref{mu_cf}, the $\Delta H$ and $\Delta S$ were obtained from thermochemical data.~\cite{themochemi_data}
The entropic terms in \eref{dS_cf} are computed from MD simulations~\cite{Berens_jcp_1983} at temperature $T$ using
\begin{eqnarray}
S = k_{\rm B} \int_0^{\infty} d\nu G(\nu) W_{\rm s}(\nu) \nonumber 
\end{eqnarray}
where,
\begin{eqnarray}
W_{\rm s} (\nu) = \frac{\beta h \nu}{ {\rm exp}(\beta h \nu) -1} - {\ln}[ 1 - {\exp}(\beta h \nu) ] \nonumber 
\end{eqnarray}
and 
\begin{eqnarray}
G(\nu) = \frac{2}{k_{\rm B}T} \sum_{J=1}^N \sum_{K=1}^3 M_J g_J^K (\nu) \nonumber  \\
\end{eqnarray}
with
\begin{eqnarray}
g_J^K (\nu) = \lim_{\tau \rightarrow \infty} \frac{1}{2\tau} \left| \int_{-\tau}^{\tau} dt  \enspace {\bf \dot R}_J^K {\rm exp}(-i2 \pi \nu t) \right|^2 \enspace . \nonumber 
\end{eqnarray}
Here $M_J$ and ${\bf \dot R}_J^K$ are the mass and $K^{th}$ velocity component of an atom $J$, respectively, and $\beta=1/(k_{\rm B} T)$. 
However, we will show later that the entropic differences in \eref{dS_cf}  can be ignored.

For the formation of Rh$_2$ cluster, 
we use 
\begin{multline} \label{chem_rxn_2}
  {\rm 2Rh(C_2H_4)_2/Y} + y{\rm H_2} \rightarrow  \\ 
  {\rm Rh_2H}_m/{\rm Y} + {\rm HY + 4C_2H_6} \enspace .
\end{multline}
%
\eref{chem_rxn_1} is not used for the formation of Rh$_2$ cluster, as the experiments were carried out using
Y--supported ${\rm Rh(C_2H_4)_2}$ clusters.~\cite{Liang_jpcc_2008} 
Computation of $\Delta G_{\rm cf}$ for \eref{chem_rxn_2} was performed in the same fashion as explained above.

A negative $\Delta G_{\rm cf}$ for the formation of a protonated Rh$_{n+1}$ cluster indicates that its formation from the parent Rh$_{n}$ cluster and Rh(C$_2$H$_4$)$_2$
(either through \eref{chem_rxn_1} or \eref{chem_rxn_2}) is thermodynamically favored.
Moreover, comparison of $\Delta G_{\rm cf}$ for Rh$_n$H$_m$ clusters with varying $m$ will help in identifying
the thermodynamically most stable protonation of the supported Rh$_n$ cluster.

In order to model the Y--zeolite 
a $2{\times}2{\times}2$ (Si$_{1536}$O$_{3072}$) supercell was taken in our calculations.
The initial coordinates were generated from the zeolite database.~\cite{Zeolite_Database}
The bulk structure was first optimized at the MM level,\cite{Sudhir_JCC_2015} 
and the optimized structure was used for QM/MM calculations.

A T25 site of zeolite (87 atoms) with adsorbate was treated using QM within a 
cubic QM box of side 23.27~{\AA}.  
One of the Si atoms in the center of the QM region was replaced by an Al atom. 
The total charge of the zeolite framework is balanced by protonating 
one of the oxygen atoms coordinated to the substituted Al.
Spin polarized calculations were carried out for supported Rh clusters. 
Multiplicity of the QM wavefunction for various 
supported Rh clusters studied here are listed in the Supporting Information. 
%
The total charge of the QM and the MM regions were always zero in these calculations.
%

%

%

For these computations Grimme's~\cite{Grimme_jcc_2006} dispersion correction was used together with the PBE density functional.
All the other computational details are the same as that in Sections~\ref{m:bench:1} and \ref{m:bench:2}.

\section{Results and Discussion}
\subsection{Testing the Implementation of CPMD/GULP QM/MM Interface}
\begin{figure}
\begin{center}
\includegraphics[width=1.00\linewidth]{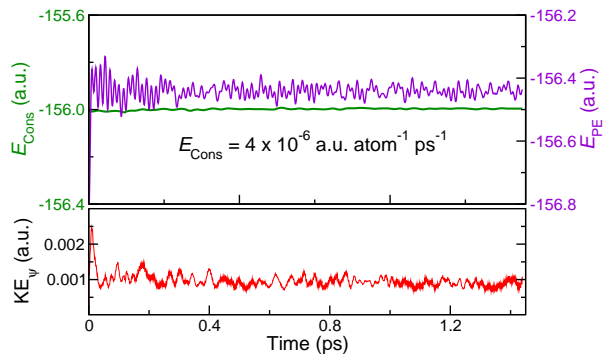}
\caption[]{
\label{E_cons_NVE_NONPOL}
Upper panel: The conserved energy $E_{\rm Cons}$ due to \eref{CP_Lg} (left axis) is plotted together with
the potential energy $E_{\rm PE}$ given by \eref{ETOT} (right axis).  Lower panel: fictitious classical orbital kinetic energy 
is plotted during the Car--Parrinello QM/MM MD simulation.
}
\end{center}
\end{figure}

To validate the implementation, Car--Parrinello MD simulation in the $NVE$ ensemble 
was carried out for the $1{\times}1{\times}1$ FAU zeolite (Si$_{192}$O$_{384}$).
Total energy conservation is a very critical test for the implementation, especially to check if the energy and the force components
from different parts of the code have been accounted correctly.
Further, a stable dynamics of orbitals can be also tested, by monitoring the fictitious orbital kinetic energy during the dynamics.
The drift in total energy was about 10$^{-6}$ a.~u. ps$^{-1}$ per atom, which is very small, indicating that the 
implementation is working correctly; see~\fref{E_cons_NVE_NONPOL}.
Also, the fictitious orbital kinetic energy fluctuation is not showing any drift (especially in the absence of a  thermostat on these degrees of freedom), 
implicating a stable Car--Parrinello dynamics. 

\begin{figure*}
\begin{center}
\includegraphics[width=1.00\linewidth]{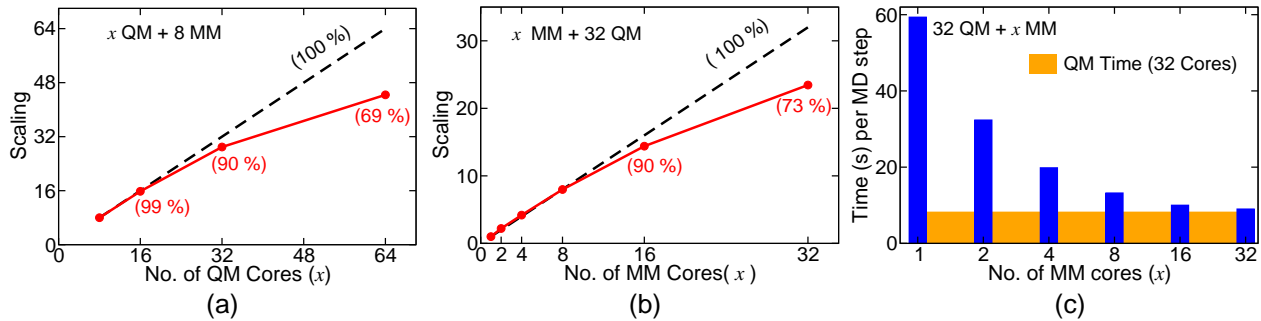}
\caption[]{
\label{scaling_fig}
Scaling performance (\eref{e:scale}) of the parallel implementation of CPMD and GULP in a QM/MM calculation is presented. In (c), blue color bars show MM clock time
and orange color shows the clock time (i.e. 9~s) using 32 QM processors.
These calculations were carried out on a computer cluster with Intel Quad--core Xeon X5560 (2.8~GHz) processors  connected using QDR (10~Gb/s) InfiniBand. 
QM and MM cores were always allocated on different compute nodes.}
\end{center}
\end{figure*}

The parallel performance of the code was analyzed based on the average clock time per MD step 
by changing the number of (computing) cores for MPI tasks allocated to CPMD and GULP. 
The clock time also included the whole I/O tasks and all the communications within and between the MM and the QM cores. 
By fixing the number of MM cores to 8, we changed the number of QM cores as $2^n$, $n=3,\cdots,6$ to verify the
CPMD scaling. 
Similarly, to analyze the GULP scaling for the same QM/MM system, we fixed the number of QM cores to 32, 
and varied the MM cores as $2^n$, $n=0,\cdots,5$.
Scaling with respect to $N_{\rm core}^0$ number of cores when $N_{\rm core}$ number of cores
was used, is given by
\begin{equation}
\label{e:scale1}
  \frac{t^0_{\rm MD}}{ t_{\rm MD} } N^0_{\rm core}
\end{equation}
where $t^0_{\rm MD}$ is the clock time per MD step for $N_{\rm core}^0$ number of cores, and
$t_{\rm MD}$ is the clock time per MD step for $N_{\rm core}$ number of cores.
Percentage of scaling is computed as
\begin{equation}
\label{e:scale}
  \frac{t^0_{\rm MD}}{ t_{\rm MD} } \frac{N^0_{\rm core}}{N_{\rm core}} \times 100 \enspace .
\end{equation}
For CPMD and GULP scaling tests, $N_{\rm core}^0$ was 8 cores and 1 core, respectively.
CPMD scales upto 69~\% (with respect to 8~cores) using 64 cores, while GULP scales 
up to 73 \% (with respect to 1~core) using 32~cores~\fref{scaling_fig}, indicating
a very good scaling performance if a large QM/MM system has to be treated.
Note that the computational time for the QM--MM electrostatics is accounted together with the CPMD time.
%
Such a dual parallelization will allow to scale down the total computational time per MD step by increasing the 
number of QM or MM cores, as appropriate for a given system and available computing resources.
This is demonstrated in \fref{scaling_fig}c where using 32 cores for CPMD and 1 core for GULP 
has a clock time of about 60~s per MD step.
Here, CPMD required only about 9~s per MD step, while GULP required 60~s per MD step,  
thus MM force calculations become the bottleneck in the QM/MM calculations using 32+1 computing cores. 
Total clock time can be systematically decreased by increasing the number of MM cores: with 32 MM cores, the clock time becomes nearly 9~s and
thus scaling the clock time per MD step of the QM/MM calculation from 60~s (with 32+1 cores) to nearly 9~s (with 32+32 cores).
\subsection{Neutral oxygen vacancy in $\alpha$--Cristobalite}

At first, we carried out $NVT$ ensemble simulation of pure $\alpha$--Cristobalite at 300~K. 
The internal structural parameters of pure SiO$_2$ from fully MM, fully QM and QM/MM MD calculations are compared with experimental values in \tref{str_MD_QM_Vs_MM}. 
The structural parameters of the inner QM regions of the QM/MM calculations are in good agreement with that of the full QM data.
Interestingly, the crystal structural data is better reproduced in fully MM calculations than in fully QM and QM/MM simulations, although the differences
are not large.
It is worth noting that near the QM/MM boundary, the structural parameters of the QM/MM atoms are
deviated more from the fully QM data; see Supporting Information.
This is expected due to the finite boundary between QM and MM regions. 
Thus, when using the QM/MM interface, we make sure that the chemically complex region is far from the boundary, 
in this case, beyond the first nearest neighboring tetrahedral site.

%
\begin{table*}
\begin{center}
\caption[]{
\label{str_MD_QM_Vs_MM}
The average value of Si--O bond length (\AA), O--Si--O and Si--O--Si angles ($^\circ$) and their standard deviation from MD at 300~K
using MM,  QM, and QM/MM potentials. The corresponding values in the crystal structure~\cite{Downs_AmMin_1994} are also listed for comparison.}
\begin{tabular}{|c|c|c|c|c|} \hline
~         &   MM   &    QM             & QM/MM (14T)       &  Expt. \\ \hline 
Si--O     &  1.60($\pm$0.03) &  1.63($\pm$0.03) & 1.62($\pm$0.03)   & 1.60(3)         \\ \hline
O--Si--O  & 109.4($\pm$2.9)  & 109.2($\pm$4.0)  & 109.4($\pm$3.8)   & 108.2--111.4   \\ \hline
Si--O--Si & 150.7($\pm$4.2)  & 142.2($\pm$6.3)  & 142.7($\pm$6.0)   & 146.4(9)       \\ \hline
\end{tabular}
\end{center}
\end{table*}
The vacancy formation energies $\Delta E_{\rm f}$ for $\alpha$--cristobalite computed from fully QM 
calculations are tabulated in \tref{data_fullQM_QMMM}. 
The converged defect formation energy is 7.49~eV.
The Si--Si distance ($r_{\rm SiSi}$) of the Si--Si bond that is formed due to the defect
is also given in \tref{data_fullQM_QMMM}, and the converged value is 2.39~{\AA}.
\begin{table} 
\begin{center}
\caption[]{
\label{data_fullQM_QMMM}
The $\Delta E_{\rm f}$ and $r_{\rm SiSi}$ obtained from fully QM and QM/MM calculations.}
\begin{tabular} {|ccc|} \hline
~         & QM & ~        \\ \hline
Supercell           & $\Delta E_{\rm f}$ (eV) & $r_{\rm SiSi}$ ({\AA}) \\ \hline
2$\times$2$\times$2 &  7.42  &  2.39  \\
3$\times$3$\times$2 &  7.49  &  2.39  \\
3$\times$3$\times$3 &  7.49  &  2.39  \\ \hline
~         & QM/MM          & ~        \\ \hline
 QM sites & $\Delta E_{\rm f}$ (eV) & $r_{\rm SiSi}$ ({\AA}) \\ \hline
 8T       & 7.69           &  2.50   \\
 14T      & 7.86           &  2.54   \\
 26T      & 7.96           &  2.58   \\ \hline
\end{tabular}
\end{center}
\end{table}
%
%
$\Delta E_{\rm f}$ and $r_{\rm SiSi}$ computed from QM/MM calculations are given in \tref{data_fullQM_QMMM} varying the size of the QM subsystem. 
Increasing the QM size does not result in a systematic convergence, which could arise in a QM/MM calculation
as the QM--MM interactions do not vary systematically with increasing QM size due to increasing boundary size (i.e. increasing number of QM--MM bond cuts).
Such a behavior is also seen in Refs.~\cite{Zipoli_jcp_2006,Peguiron_JCP_2015} 
%
%
Thus, for e.g. taking energy differences between QM/MM systems of different QM sizes shall be done with care.
It may be noted that $r_{\rm SiSi}$ is overestimated in QM/MM geometry optimizations which could be due to fixed
 Si$_{\rm QM}$--O$_{\rm QM}$--H$_{\rm capH}$--Si$_{\rm MM}$ boundary atoms and MM atoms in these calculations to speed up the convergence. 
In the following we have computed the ensemble average of $r_{\rm SiSi}$ from MD simulations where all the 
atoms are relaxed and the agreement with fully QM data is improved.

\begin{figure}
\begin{center}
\includegraphics[width=1.00\linewidth]{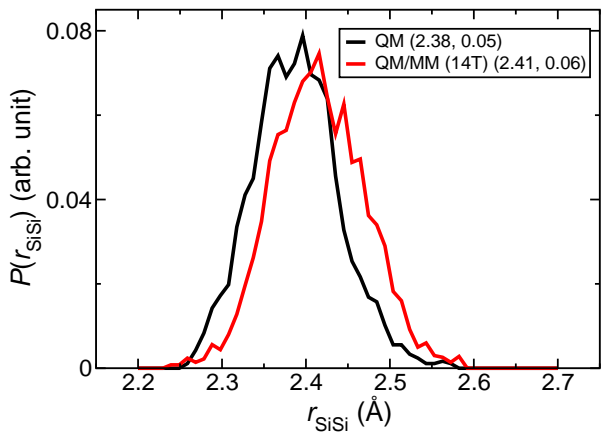}
\caption[]{
\label{r_SiSi_MD}
Probability distribution of $r_{\rm SiSi}$ from QM/MM (14~T) and QM MD simulations at 300~K. The average bond length and standard deviation (in {\AA}) 
are given in parenthesis.
}
\end{center}
\end{figure}

As next, we carried out $NVT$ ensemble simulations at 300~K for a QM (3$\times3\times3$ supercell) 
and QM/MM calculations (with 14~T QM size).
%
Distribution of $r_{\rm SiSi}$  obtained form these simulations is given in \fref{r_SiSi_MD}), where it is
observed that the average and the standard deviation of the QM and the QM/MM distributions differ only by 
0.03~{\AA}, and  0.01~{\AA}, respectively. 
%
%
%
%
%

\begin{figure}
\begin{center}
\includegraphics[width=1.00\linewidth,angle=0]{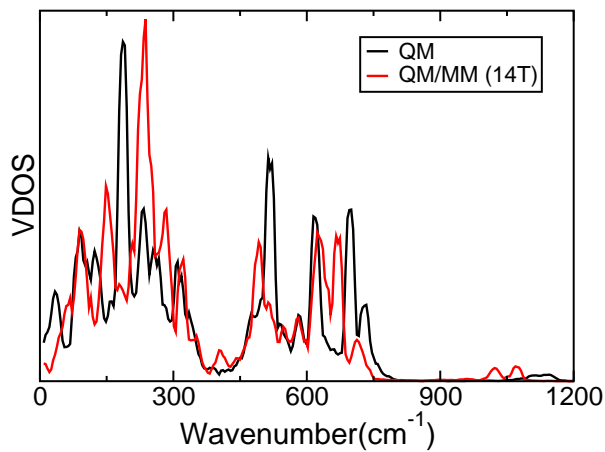} 
\caption[]{
\label{vdos_MD}
The VDOS of the 
Si-Si
unit computed from a full QM and QM/MM (with QM size 14T) MD simulations. 
A frequency window of 20 cm$^{-1}$ was used for smoothening the spectra. 
}
\end{center}
\end{figure}

From fully QM and QM/MM MD simulations, vibrational density of states (VDOS) was computed from the Fourier transform of 
the autocorrelation for the Si--Si bond distance; see~\fref{vdos_MD}. 
Main features of the VDOS are reproduced in QM/MM calculations. 
The full VDOS of O$_3$Si--SiO$_3$ unit also agrees qualitatively well within 50~cm$^{-1}$ (see Supporting Information).

%

\begin{figure}
\begin{center}
\includegraphics[width=1.00\linewidth,angle=0]{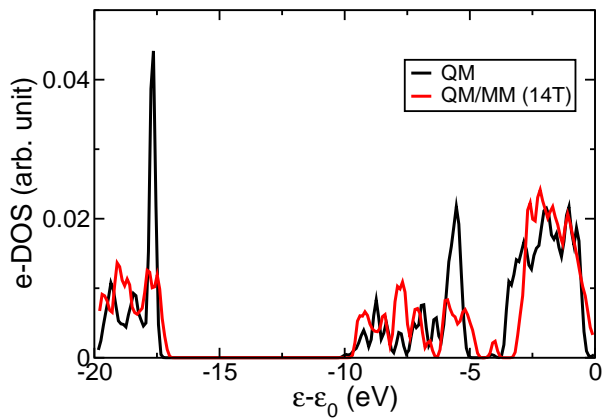}
\caption[]{
\label{total_DOS}
e-DOS computed from both fully--QM and QM/MM calculations. $\epsilon$ is the orbital energy and $\epsilon_{0}$ is the orbital energy of 
highest occupied state.
}
\end{center}
\end{figure}

We have also compared the total density of the Kohn-Sham electronic states (e-DOS) of all the occupied molecular orbitals; see \fref{total_DOS}.
There is a qualitative agreement between the two data, especially the valence states, reflecting the reliability of the implementation. 
We do not notice the presence of any arbitrary electronic states, which would be otherwise noticeable in the e-DOS. 
A decreased relative density for some of the semi-core states could be due to small QM size compared to the fully QM calculation.
\subsection{Study of Y--Zeolite Supported Rh$_n$ ($n=2-6$) Clusters}
Here we investigate the stable structures of hydrogenated Rh$_n$ clusters on Y--zeolite, and  their
 formation free energies are computed as a function of $T$ and at some fixed $p$.
%

\begin{figure}
\begin{center}
\includegraphics[width=1.00\linewidth]{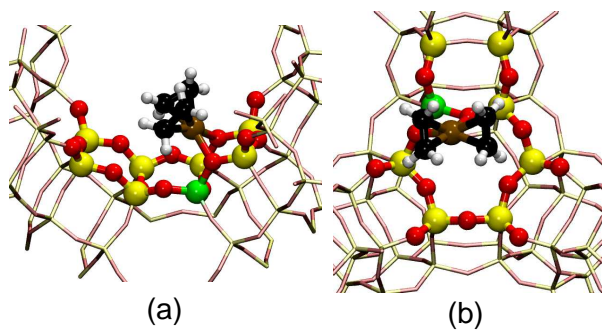} 
\caption[]{
\label{Rh1Et2_HY_fig}

Optimized structures of Y--zeolite supported Rh(C$_2$H$_4$)$_2$ at the (a) 4R and (b) 6R adsorption sites. 
Atom color code: H--white, C--black, Al--green and Rh--ochre.
}
\end{center}
\end{figure}

%
The structure of Rh(C$_2$H$_4$)$_2$/Y at two different adsorption sites 4R and 6R (see~\fref{Rh1Et2_HY_fig}) 
were first analyzed and compared with the available experimental data.
The Rh atom coordinates to two of the zeolite O atoms and both the ethylene ligands are $\pi$-bonded to the metal atom in the adsorbed structure.
We find that metal complex is more stable at the 4R site by 14~kJ~mol$^{-1}$ than at the 6R site.
The interatomic distances 
of the optimized structure are given in \tref{Rh1Et2Y}.
This structural data agrees reasonably well with the experimental data.~\cite{Ehresmann_angewie_2006}
Interestingly, the 6R adsorption structure agrees better with the experimental data compared to the 4R adsorption structure.
%
%
\begin{table}
\begin{center}
\caption[]{
\label{Rh1Et2Y}
Structural parameters of the optimized structure of Y--zeolite supported Rh(C$_2$H$_4$)$_2$  at  4R and 6R
adsorption sites. Relative energy (with respect to adsorption at 4R site) is in kJ~mol$^{-1}$ and 
the interatomic distances are in {\AA}. Experimental structural parameters\cite{Ehresmann_angewie_2006} are also 
listed here for comparison.}
\begin{tabular}{|c|c|c|c|} \hline
        ~                 & ${\Delta}E$ & Rh--O &  Rh--C \\ \hline
{4R} &  0       & 2.22     & 2.12  \\ \hline
{6R} &  14      & 2.18    & 2.10  \\ \hline
{Expt.} &  --   & 2.19    & 2.09  \\ \hline 
\end{tabular}
\end{center}
\end{table}

Now we compute the $\Delta G_{\rm cf}$ for Rh$_n$H$_m$/Y, $n=2-6$, based on which we identify the most stable protonated structure of a cluster adsorbed within Y--zeolite.
The vibrational entropy, $S$ at 300~K of zeolite--HY 
nearly converges to
12.70~${\rm J~mol^{-1}~K^{-1}}$ after 6~ps of the simulation; see Supporting Information. 
The converged entropies of Rh(C$_2$H$_4$)$_2$/Y, Rh$_3$H$_7$/Y and Rh$_4$H$_9$/Y are given in ~\tref{S_table}. 
Thus, we find that the entropy contribution ($T\Delta S_{\rm cf}(T)$) 
to the $\Delta G_{\rm cf}$ of formation Rh$_4$H$_9$/Y from Rh$_3$H$_7$/Y is negligible.
Thus, we ignore the entropic terms in computing $\Delta G_{\rm cf}$ for other supported clusters in this study.

\begin{figure*}
\begin{center}
\includegraphics[width=0.80\linewidth]{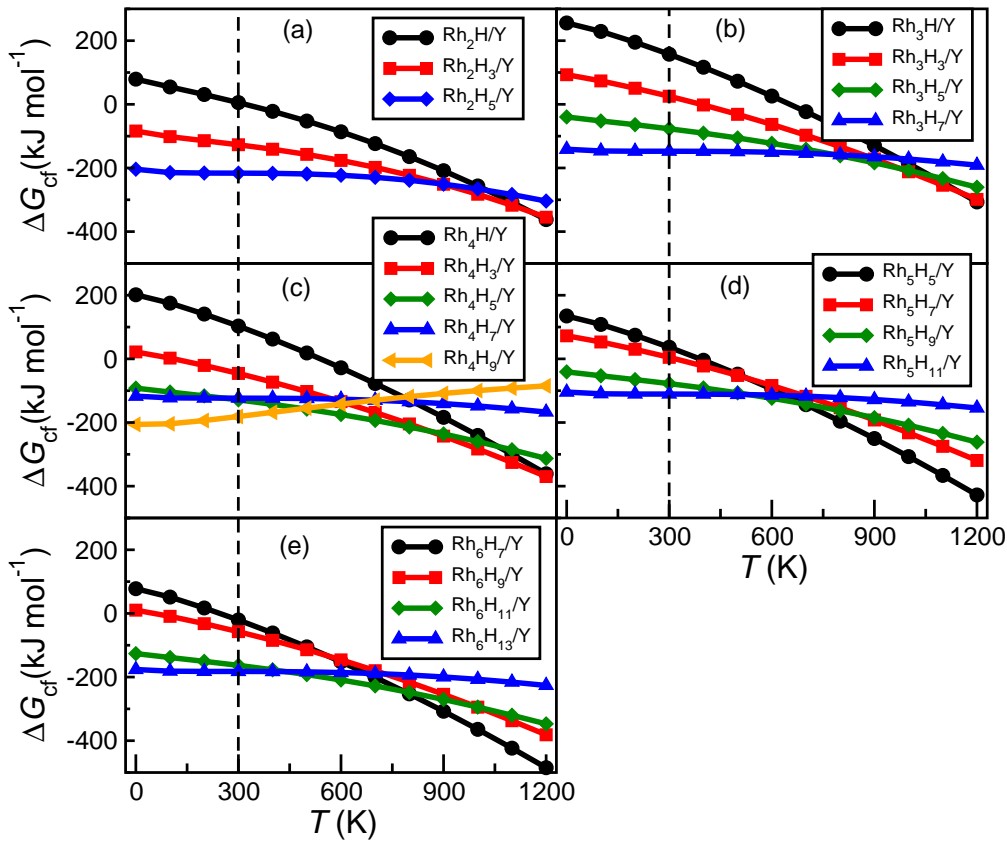}
\caption[]{
\label{Cluster_form_Rh2-6}
The $\Delta G_{\rm cf}$ for Rh$_n$H$_m$/Y ($n$=2--6; $m=1, \cdots ,2n+1$) cluster as a
function of $T$ and at $p_{\rm H_2}$=$p_{\rm C_2H_6}$=1.0 bar pressure.
}
\end{center}
\end{figure*}

The $\Delta G_{\rm cf}$ of formation of Rh$_2$H$_m$/Y ($m$=1, 3, and, 5) cluster as 
a function of $T$ at $p_{\rm H_2}$=$p_{\rm C_2H_6}$=1.0 bar is given in ~\fref{Cluster_form_Rh2-6}a. 
At 300~K and $p_{\rm H_2}$=$p_{\rm C_2H_6}$=1.0 bar pressure,  thermodynamically the most stable cluster 
is Rh$_2$H$_5$ (i.e. with the lowest $\Delta G_{\rm cf}$ value) for $n=2$.
At higher $T$, the clusters with less number of H atoms become more stabilized, as shown in \fref{Cluster_form_Rh2-6}a. 
Similar studies were also carried out for Rh$_n$H$_m$ for $n=3-6$, and same conclusions can be drawn from these data (\fref{Cluster_form_Rh2-6}b-e). 
 For all the cases,  the thermodynamically stable state at 300~K and  $p_{\rm H_2}$=$p_{\rm C_2H_6}$=1.0 bar pressure, has $m=2n+1$.
%
%
Another crucial information from these plots is that  $\Delta G_{\rm cf}$ values of all the clusters are negative at ambient condition, 
showing that these clusters can spontaneously form from their parent clusters (at the thermodynamic limit).
%

%
%

\begin{figure*}
\begin{center}
\includegraphics[width=1.00\linewidth]{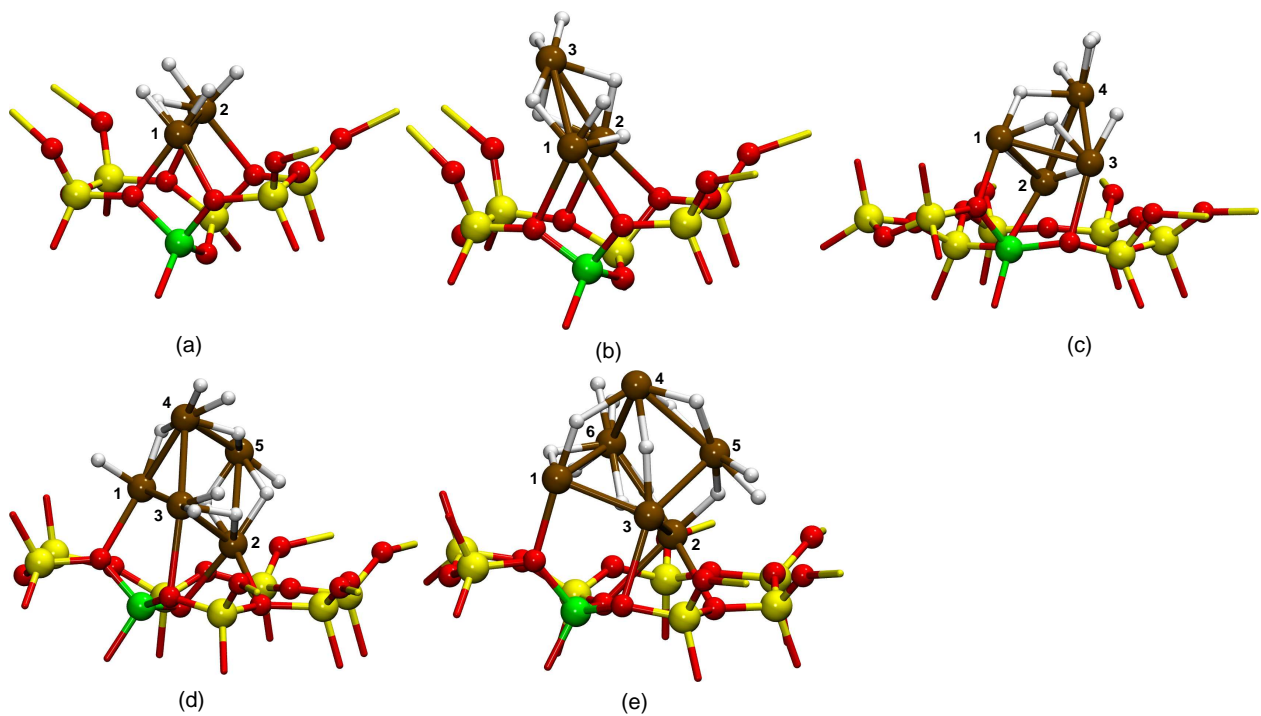}
\caption[]{
\label{Rh_Str_Rh2-6_MD}
Snapshots of Y--zeolite supported Rh$_n$H$_m$ clusters from canonical MD simulations at 300~K: (a) Rh$_2$H$_5$  (4R); (b) Rh$_3$H$_7$  (4R) ; 
(c) Rh$_4$H$_9$  (6R); (d) Rh$_5$H$_{11}$  (6R); (e) Rh$_6$H$_{13}$  (6R). Only a few atoms near the adsorption site are shown here. Labels of Rh atoms are
also shown.}
\end{center}
\end{figure*}

%
The equilibrium structural details
and relative energies of various structures of Rh$_n$H$_{2n+1}$ clusters adsorbed at 4R and 6R sites 
are given in the~\tref{RhnHm_Str}. 
%
%
Minimum energy configurations are  displayed in \fref{Rh_Str_Rh2-6_MD}; see also Supporting Information.
The Rh$_2$H$_5$ and Rh$_3$H$_7$ clusters are more stable at 4R sites than 6R sites, 
whereas the Rh$_4$H$_9$, Rh$_5$H$_{11}$ and Rh$_6$H$_{13}$ are more stable in 6R sites.
\begin{table*}
\begin{center}
\setlength{\tabcolsep}{0.15cm}
\caption[]{
\label{RhnHm_Str}
Structure of various Y--zeolite supported Rh$_n$H$_m$ clusters at 4R and 6R sites.
Average distances are given, 
when more than one such distance is present within the molecule. 
The relative energy $\Delta E$ (in kJ~mol$^{-1}$) is with respect to 4R  and the interatomic distances are 
in {\AA} unit.
%
$n\mathrm H_{\rm brid}$ and $n\mathrm H_{\rm term}$ are the number of bridging and terminal H atoms, respectively.} 
\begin{tabular}{|c|c|c|c|c|c|c|} \hline
Structure      & Site & $\Delta E$ & Rh--Rh & Rh--O & $n\mathrm H_{\rm brid}$ & $n \mathrm H_{\rm term}$ \\ \hline
Rh$_2$H$_5$    & 4R & 0.0  & 2.59   & 2.31  & 1 & 4 \\ 
               & 6R & 1.8  & 2.59   & 2.25  & 1 & 4 \\ \hline
Rh$_3$H$_7$    & 4R & 0.0  & 2.71   & 2.26  & 4 & 3 \\ 
               & 6R & 22.7 & 2.70   & 2.32  & 3 & 4 \\ \hline
Rh$_4$H$_9$    & 4R & 0.0  & 2.82   & 2.36  & 6 & 3  \\ 
               & 6R &-44.2 & 2.84   & 2.32  & 6 & 3  \\  \hline
Rh$_5$H$_{11}$ & 4R & 0.0  & 2.73   & 2.30  & 5 & 6  \\ 
               & 6R &-27.9 & 2.82   & 2.22  & 7 & 4  \\ \hline
Rh$_6$H$_{13}$ & 4R & 0.0  & 2.88   & 2.31  & 8 & 5  \\ 
               & 6R &-41.2 & 2.90   & 2.26  & 8 & 5  \\ \hline
\end{tabular}
\end{center}
\end{table*}
%
The average Rh--Rh distances increase with cluster size and number of hydrogen atoms. 
%
The average Rh--O distances vary 2.22--2.35 {\AA}, and 
no systematic variation was observed with increasing cluster size.
Overall, our prediction of the protonation states of the supported Rh$_n$ clusters also agree with
the result of Markova et.~al.~\cite{Markova_jpcc_2015}
%
%

\begin{figure*}
\begin{center}
\includegraphics[width=1.00\linewidth]{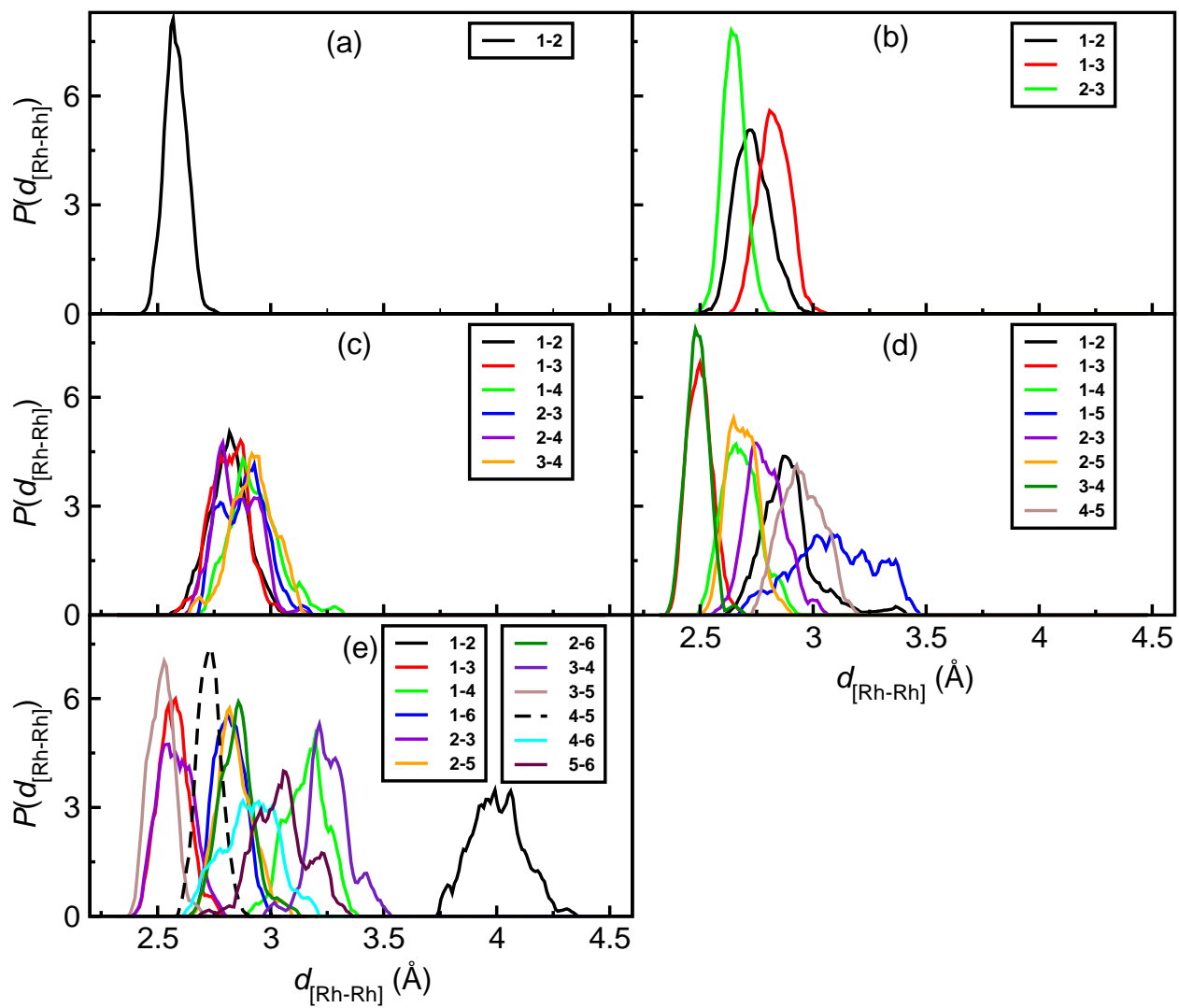}
\caption[]{
\label{Rh_bond_dist_all}
 Distribution of various Rh--Rh bond distances in a Rh$_n$H$_{2n+1}$ cluster from a canonical ensemble MD simulation at 300~K: 
 (a) Rh$_2$H$_5$ (4R) ; (b) Rh$_3$H$_7$ (4R); (c) Rh$_4$H$_9$ (6R); (d) Rh$_5$H$_{11}$  (6R); (e) Rh$_6$H$_{13}$  (6R).
Atom labels are as shown in \fref{Rh_Str_Rh2-6_MD}. } 
\end{center}
\end{figure*}

Finally, we analyze the structure of the supported clusters from QM/MM MD simulation at 300~K.
The Rh--Rh bond distance distribution from these simulations for Rh$_n$H$_{2n+1}$, $n=2-6$ are given
in \fref{Rh_bond_dist_all}. 
 The protonated Rh$_2$ cluster is coordinated to four support oxygen atoms, with two Rh--O bonds per Rh atom; see \fref{Rh_Str_Rh2-6_MD}.
Rh$_3$ forms a distorted triangle, with two Rh atoms directly interacting with the support oxygen atoms similar to the case of Rh$_2$.  
 All the three Rh--Rh bond distributions have different averages.
 Rh1--Rh3 bond is weaker than the other two bonds, because both Rh1 and Rh3 have more terminal hydrogen atoms compared to Rh2.
The protonated Rh$_4$ cluster, which preferentially adsorb at the 6R site, has a distorted tetrahedral structure with three Rh atoms directly coordinating
to  support oxygen atoms.
The bond distributions functions are largely localized between 2.5--3.0~{\AA}, and are broader than the Rh$_3$ case.
Especially, the some of the bonds made by Rh4 atom become nearly dissociated ($\sim 2.9$~{\AA}) as it is coordinated to more number of hydrogen atoms.
A highly distorted pyramidal structure was observed for the protonated Rh$_5$ cluster at the preferred 6R site.
Similar to Rh$_4$, three Rh atoms are directly coordinating to support oxygen atoms.
One of the bonds, Rh1--Rh5, was mostly found broken during the simulation, which is obvious from the broadness of its distribution and its average going beyond 3~{\AA}.
The Rh3--Rh4 and Rh3--Rh1 bonds in Rh$_5$ cluster are more stronger than the other Rh--Rh bonds, 
as clear from their distributions which are having smaller standard deviations and smaller average values compared to other bonds.
We ascribe this to the fact that more number of hydrogen atoms are coordinated to Rh3 atom compared to other Rh atoms in the cluster.
Rh$_6$ is adsorbed at the 6R site, and has a distorted octahedral structure.
The bond length distributions of Rh$_6$ cluster are more complex.
Clearly, no covalent bond exists between Rh1--Rh2, Rh2--Rh3, and Rh1--Rh4 as their bond distances are lying above 3~{\AA}. 
Rh5--Rh6, and Rh4--Rh6 bonds were broken and formed numerous times during the MD simulation. 
%

%
\begin{table}
\begin{center}
\caption[]{
\label{S_table}
Vibrational entropy, $S$, computed at 300~K from constant temperature MD simulations for various systems. 
}
\begin{tabular}{|c|c|} \hline
~                    & $S({\rm J mol^{-1} K^{-1}})$  \\  \hline 
HY                   & 12.70                          \\  \hline 
Rh(C$_2$H$_4$)$_2$/Y & 12.90  \\ \hline
Rh$_3$H$_7$/Y        & 12.84  \\ \hline
Rh$_4$H$_9$/Y        & 12.95  \\ \hline
\end{tabular}
\end{center}
\end{table}
%


%
%
%

\section{Conclusions}
We have reported the development of CPMD/GULP interface for hybrid QM/MM MD simulation of periodic solids and surfaces.
The whole QM and MM subsystems can be relaxed during geometry optimizations and MD simulations.
The developed program is shown to have good total energy conservation during MD simulations and stable fictitious dynamics of orbital degrees of freedom
were observed during a microcanonical ensemble Car--Parrinello MD simulation of a Y--zeolite system, manifesting the correctness of the implementation.
Both CPMD and GULP programs can carry out force--evaluations in parallel under the same MPI environment, and is shown to be beneficial 
for treating large systems.
%
%

The QM/MM interface program was further tested by studying the  neutral oxygen vacancy in $\alpha$-cristobalite.
A reasonable agreement of the computed structural, dynamic and electronic properties from a fully QM simulation
and QM/MM simulation was observed.
%
The vacancy formation energy is also reproduced well, although a systematic convergence with increasing QM size was not observed.

Structure of protonated Rh$_n$ clusters supported in Y--zeolite was subsequently investigated.
Free energy of formation of Y--zeolite supported protonated Rh$_n$ clusters from protonated Rh$_{n-1}$ and Rh$_2$(C$_2$H$_4$)$_2$ 
at hydrogen atmosphere was computed for various temperature and at given partial pressures of H$_2$ and C$_2$H$_6$.
We find that Rh$_n$H$_m$ clusters with $m=2n+1$ are favorable at low $T$ than $m<2n+1$, while the clusters with less number of H atoms become
 increasingly stabilized with increase in $T$.
 Our results are in qualitative agreement with a previous theoretical work.
The thermodynamically stable Y--zeolite supported Rh$_n$H$_{2n+1}$  clusters at 300~K was further 
studied using constant temperature MD simulation at 300~K and their structural properties are reported here.

The developed CPMD/GULP QM/MM interface program is suited for studying complex catalytic reactions by carrying out enhanced sampling MD simulations.
In particular, the interface allows us to perform Car--Parrinello based metadynamics simulations of catalytic reactions, which is advantageous in studying supported metal catalysis.
Such applications using the QM/MM interface are ongoing in our laboratory.

{\bf Acknowledgments} \\
Authors acknowledge the HPC facility at IIT Kanpur.
SKS thanks CSIR and IIT Kanpur for his Ph.D. fellowship. \\
%

%
\clearpage
{\bf TOC Figure}  \\
\begin{figure} [h]
\begin{center}
\includegraphics[width=1.00\linewidth]{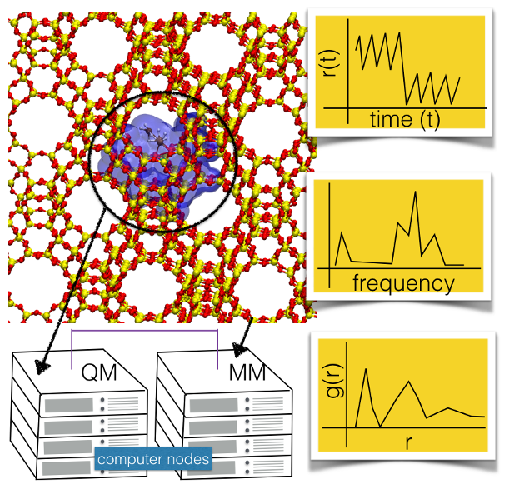} 
\end{center}
\end{figure} \\
{\bf TOC Caption} \\
A highly parallel CPMD/GULP QM/MM Interface is developed here for
performing molecular dynamics simulations of periodic solids. Application
of this code for studying static and dynamic properties of zeolite
supported Rh clusters is also presented.
\end{document}